\documentclass[twocolumn,showpacs,aps,prl,letter,amsmath,amssymb,superscriptaddress]{revtex4-1}
\usepackage{bm,color,bbm}
\usepackage{mathtools,graphicx,natbib}

\usepackage[english]{babel}
\usepackage[utf8x]{inputenc}
\usepackage[T1]{fontenc}
\usepackage{float}

\newcommand{\beq}{\begin{equation}}
\newcommand{\eeq}{\end{equation}}
\newcommand{\bqa}{\begin{eqnarray}}
\newcommand{\eqa}{\end{eqnarray}}
\newcommand{\nn}{\nonumber}

\newcommand{\smallfrac}[2]{\mbox{$\frac{#1}{#2}$}}

\newcommand{\ket}[1]{ |{#1} \rangle}

\newcommand{\half}{\smallfrac{1}{2}}

\newcommand{\sq}[1]{\left[ {#1} \right]}

\newcommand{\tr}[1]{{\rm Tr}\sq{ {#1} }}

\newcommand{\blk}{\color{black}}
\newcommand{\blu}{\color{black}}
\definecolor{maroon}{rgb}{0.7,0,0}

\definecolor{ngreen}{rgb}{0.3,0.7,0.3}

\definecolor{golden}{rgb}{0.8,0.6,0.1}

\usepackage{amsmath}
\usepackage{graphicx}
\usepackage{epstopdf}
\usepackage[colorinlistoftodos]{todonotes}

\begin{document}

\title{Strong unitary and overlap uncertainty relations: theory and experiment}

\author{Kok-Wei Bong}
\affiliation{Centre for Quantum Computation and Communication Technology (Australian Research Council), Centre for Quantum Dynamics, Griffith University, Brisbane, QLD 4111, Australia.}

\author{Nora Tischler}
\affiliation{Centre for Quantum Computation and Communication Technology (Australian Research Council), Centre for Quantum Dynamics, Griffith University, Brisbane, QLD 4111, Australia.}

\author{Raj B. Patel}
\affiliation{Centre for Quantum Computation and Communication Technology (Australian Research Council), Centre for Quantum Dynamics, Griffith University, Brisbane, QLD 4111, Australia.}

\author{Sabine Wollmann}
\affiliation{Centre for Quantum Computation and Communication Technology (Australian Research Council), Centre for Quantum Dynamics, Griffith University, Brisbane, QLD 4111, Australia.}
\affiliation{Quantum Engineering Technology Labs, H. H. Wills Physics Laboratory and Department of Electrical \& Electronic Engineering, University of Bristol, BS8 1FD, UK.
}

\author{Geoff J. Pryde}
\email{g.pryde@griffith.edu.au}
\affiliation{Centre for Quantum Computation and Communication Technology (Australian Research Council), Centre for Quantum Dynamics, Griffith University, Brisbane, QLD 4111, Australia.}

\author{Michael J. W. Hall}

\affiliation{Centre for Quantum Computation and Communication Technology (Australian Research Council), Centre for Quantum Dynamics, Griffith University, Brisbane, QLD 4111, Australia.}
\affiliation{Department of Theoretical Physics, Research School of Physics and Engineering, Australian National University, Canberra ACT 0200, Australia}

\begin{abstract}
We \blk derive \blk and experimentally investigate a  strong  uncertainty relation valid for any $n$ unitary operators, which  implies the   standard  uncertainty relation \blk and others \blk as special cases,   and which can be written in terms of   geometric phases. It is saturated by every pure state of any $n$-dimensional quantum system, generates a tight overlap uncertainty relation for the transition probabilities of any $n+1$ pure states, and gives an upper bound for the out-of-time-order correlation function.  We test these uncertainty relations experimentally for photonic polarisation qubits,  including the minimum uncertainty states of the overlap uncertainty relation, via interferometric measurements of generalised  geometric  phases.
\end{abstract}

\maketitle

\paragraph{Introduction.---} 

 \blk Uncertainty relations are one of the most important foundations of physics, defining the limits on what is possible in a quantum world. Their implications range from bounds in quantum metrology~\cite{braun,met}, through the security of quantum cryptographic schemes~\cite{berta,tom}, and to the measurement and control of deeply quantum systems~\cite{wollman}. We present a very powerful, yet simple, uncertainty relation for the reversible transformations of a quantum system (represented by unitary operators), which unifies, generalises, and significantly strengthens previous results. For example, our unitary  uncertainty relation for $n$ operators: (i)~is saturated by {\it every} pure state of an $n$-dimensional Hilbert space (and by all pure qubit states); (ii)~is stronger than, and can be used to derive, the standard Heisenberg and Robertson-Schr\"odinger uncertainty relations~\cite{rob,schr}, and various others in the literature~\cite{jed,li,branciard,pati,bagchi,rudnicki}; (iii)~leads to an upper bound for the out-of-time-order correlator---of strong interest in quantum thermalisation, chaos and information scrambling, for both many-body and black-hole physics~\cite{hosur, mald, swingle,otocexp1,otocexp2,otocexp3}; and (iv)~generates a strong inequality for the transition probabilities connecting any $n+1$ pure states. 
Our relation can therefore be viewed as an `ur'-uncertainty relation which unifies a number of seemingly disparate quantum concepts. We experimentally investigate this uncertainty relation, and its implications for transition probabilities,  via  robust interferometric measurements  of generalised  geometric  phases~\cite{panch,barg,berry, aharonov, mukunda} on polarisation qubits (extendable to any $n$ unitaries). \blk
\begin{figure*}[!t]
	\centering
	\includegraphics[width=.87\textwidth]{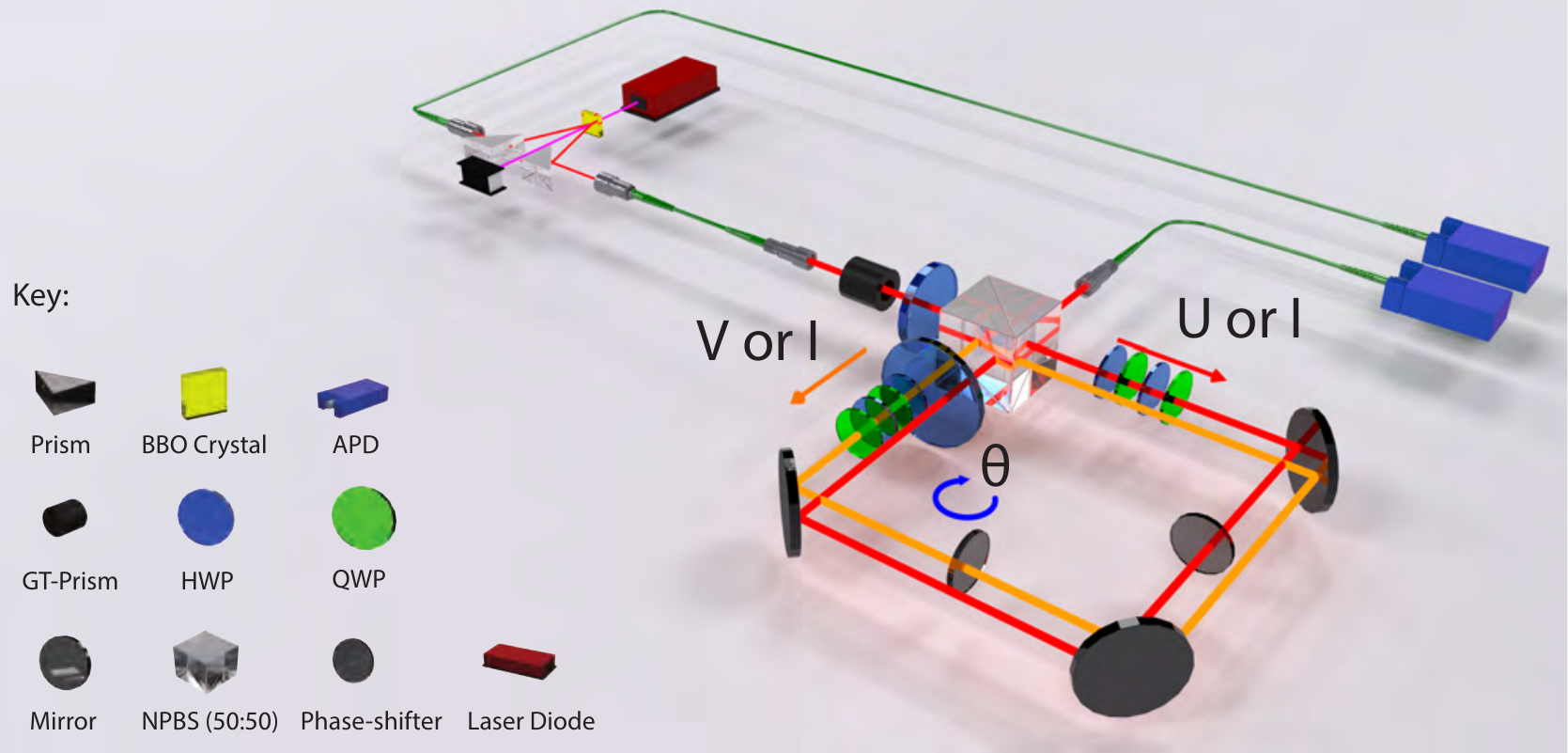}
	\caption{  Experimental setup for testing the unitary and overlap uncertainty relations. Pairs of single photons are generated via SPDC using a type-I BBO crystal. The signal photon is prepared in an arbitrary linear polarisation state using a Glan-Taylor (GT) prism followed by a half-waveplate (HWP). After entering a displaced Sagnac interferometer at a 50:50 non-polarising beamsplitter (NPBS), the photon traverses the interferometer in a superposition of the transmitted (red) and reflected (orange) paths. Unitary operators $U$, $V$ and $I$ are implemented using HWPs and quarter-waveplates (QWP). An additional HWP compensates for the birefringent phase upon reflection at the NPBS. A glass element in one path is tilted to act as a phase-shifter, while a fixed element in the other path keeps the path-length difference to within the coherence length. Two avalanche photodiodes (APDs) detect the signal and idler (heralding) photons. }
	\label{fig:expt_setup}
\end{figure*}

\paragraph{Unitary uncertainty relation.---}

For $n$ unitary operators $U_1, U_2, \dots, U_n$ and quantum state $\rho$, define $U_0=I$ and $v^{(j)}=U_j\rho^{1/2}$. For any given set of $n+1$ vectors $\{v^{(j)}\}$ with inner product $(\cdot,\cdot)$, the corresponding Gram matrix $G$, with coefficients $G_{jk} = (v^{(j)},v^{(k)})$,  is positive semidefinite~\cite{gram}. 
Hence, choosing the inner product $(A,B)=\tr{A^\dagger B}$  on the vector space of linear operators,  one has the unitary uncertainty relation (UUR)
\beq \label{uur}
\det G\geq 0,\qquad G_{jk} := \tr{\rho \,U_j^\dagger U_k} = \langle U^\dagger_jU_k\rangle 
\eeq
\blu (and, more generally, the stronger relation $G\geq 0$). \blk We note that a similar method was used  by Robertson to obtain an uncertainty relation for $n$ Hermitian operators and a pure state~\cite{rob34}. For $n=2$ the UUR reduces to
\beq \label{uv}
{\rm Var}\,U\,{\rm Var}\,V \geq |\langle U^\dagger V\rangle - \langle U^\dagger\rangle\langle V\rangle|^2 
\eeq
 for two unitary operators $U$ and $V$, recently obtained elsewhere by less simple means~\cite{pati,bagchi}, where the variance of unitary operator $U$ is defined by ${\rm Var}\,U:=1-|\langle U\rangle|^2$. 
 \blu  As well as a direct measure of uncertainty, vanishing only for eigenstates of $U$~[27], ${\rm Var}\,U$ quantifies the disturbance of pure states by $U$: it reaches its minimum value of 0 for a nondisturbing rephasing of the state,  $U|\psi \rangle=\mathrm{exp}(i \theta)|\phi\rangle$, and its maximum value of 1 for the maximally disturbing case that $U$ transforms $|\psi\rangle$ to an orthogonal state. \blk 

 The $n=3$ case is discussed in the Supplemental Material~\cite{sm}. It is further shown there that expanding $U=e^{i\epsilon A}, V=e^{i\epsilon B}$  in $\epsilon$ in Eq.~(\ref{uv}) yields the standard Robertson-Schr\"odinger uncertainty relation~\cite{rob,schr}
\beq  \label{rs}
{\rm Var}A\,{\rm Var}B \geq \frac{1}{4}|\langle [ A, B]\rangle|^2 + {\rm Cov}(A,B)^2,
\eeq
for two observables represented by Hermitian operators $A$ and $B$ (with the quantum covariance defined by
${\rm Cov}(A,B) := \half \langle AB+BA\rangle - \langle A\rangle\langle B\rangle$), and that several recent uncertainty relations~\cite{jed,li,branciard,rudnicki} can also be obtained from the UUR in Eq.~(\ref{uur}).

\blu To determine the states that saturate the UUR, \blu i.e., its minimum uncertainty states, \blk note that the determinant of a Gram matrix vanishes if and only if the vectors $v^{(j)}$ are linearly dependent~\cite{gram}. For a pure state $\rho=|\psi\rangle\langle\psi|$ this is equivalent to linear dependence of  $|\psi\rangle,U_1|\psi\rangle,\dots,U_n|\psi\rangle$, which is always satisfied for the case of a Hilbert space with dimension $d\leq n$. Hence, {\it every} pure state is a minimum uncertainty state for this case, emphasising the strength of the UUR. In particular, Eq.~(\ref{uur}), and hence Eqs.~(\ref{uv}) and (\ref{rs}), are saturated by all qubit pure states. Conversely, a mixed qubit state is a minimum uncertainty state of Eq.~(\ref{uv}) if and only if $[U,V]=0$, i.e., if and only if $U$ and $V$ correspond to rotations about the same axis of the Bloch sphere (see Supplemental Material \cite{sm}).

The UUR is invariant under $U_j\rightarrow e^{i\phi_j}U_j$ 
 (even though $G_{jk}$ in Eq.~(\ref{uur}) is not), i.e., under physically equivalent unitaries~\cite{sm}. 
Indeed, for a pure state $|\psi\rangle$, Eq.~(\ref{uur}) can be  rewritten   in terms of the Bargmann projective invariants $B_{j_1\dots j_r}:=\tr{|\psi_{j_1}\rangle\langle\psi_{j_1}|\dots |\psi_{j_r}\rangle\langle\psi_{j_r}|}$, \blu invariant under rephasings $|\psi_{j_r}\rangle\rightarrow e^{i\phi_{j_r}}|\psi_{j_r}\rangle$, \blk where $|\psi_{j+1}\rangle:=U_j|\psi\rangle$~\cite{barg,sm} (these invariants are closely related to  geometric phases~\cite{mukunda}). For example, Eq.~(\ref{uv}) becomes
\beq \label{barg}
\cos\Phi \geq \frac{T_{12}+T_{13}+T_{23} -1}{2\sqrt{T_{12}T_{13}T_{23}}} ,
\eeq
where $T_{jk}=|\langle\psi_j|\psi_k\rangle|^2=B_{jk}$ is the transition probability between $|\psi_j\rangle$ and $|\psi_k\rangle$, and $\Phi$ is the  phase of \blu the complex number \blk $B_{123}$. The saturation of this inequality for all pure qubit states corresponds to an identity in spherical trigonometry~\cite{trig, sm}.  For  general  mixed states, the UUR can be tested via the measurement of  suitably  generalised Bargmann invariants, as reported below. In particular, Eq.~(\ref{uv}) is equivalent to
\beq \label{bargen}
\cos\Phi \geq \frac{|\langle U\rangle|^2+|\langle V\rangle|^2+ |\langle U^\dagger V\rangle|^2 -1}{2|\langle U\rangle\langle U^\dagger V\rangle\langle  V^\dagger\rangle|} ,
\eeq
generalising Eq.(\ref{barg}), where $\Phi$ is the phase of the generalised Bargmann invariant $\langle U\rangle\langle U^\dagger V\rangle\langle V^\dagger \rangle$~\cite{sm}. 

\paragraph{Overlap uncertainty relation.---}

Overlap uncertainty relations reflect the nonclassical property that even pure quantum states typically overlap, important for quantum state discrimination and quantum metrology~\cite{braun,met,barnett,trace}, and in SWAP-tests \cite{rbpFredkin} for quantum communication \cite{Buhrman}.  For example, the overlap between two phase-shifted optical modes $|\psi\rangle$ and $e^{-iN\chi}|\psi\rangle$ is $T_\chi=|\langle\psi|e^{-iN\chi}|\psi\rangle|^2= 1 -\chi^2(\Delta N)^2+O(\chi^4)$, implying that a small overlap $T_\chi\ll1$, as required to resolve a small phase shift~$\chi$, requires a large photon number uncertainty $\Delta N\gtrsim 1/\chi$. Our unitary uncertainty relation unifies quantum limits on state preparation and overlap, by generating a tight  overlap  uncertainty relation for any given set of $n+1$ pure states.   

For example, noting  that $\cos\Phi\leq1$, Eq.~(\ref{barg}) immediately yields the overlap uncertainty relation (OUR)
\beq \label{our}
T_{12}+T_{13}+T_{23} - 2 \sqrt{T_{12}T_{13}T_{23}} \leq 1 .
\eeq 
for the transition probabilities connecting any three pure states. This relation is tight, being saturated if and only if the states lie on a geodesic in Hilbert space, and for qubits corresponds to their Bloch vectors lying on a great circle~\cite{sm}.  It is also a very strong constraint---stronger, e.g., than the  overlap uncertainty relation 
$\sqrt{1-T_{12}}  \leq  \sqrt{1-T_{13}} + \sqrt{1-T_{23}}$,
for the transition probabilities of any three pure states $|\psi_1\rangle, |\psi_2\rangle,|\psi_3\rangle$, corresponding to the triangle inequality for trace distance~\cite{trace,sm}. For states $|\psi\rangle$, $U|\psi\rangle$, $V|\psi\rangle$, with fixed $U$ and $V$,   saturation of the OUR   determines the corresponding minimum uncertainty states $\{|\psi\rangle\}$, as investigated experimentally below.  More generally, the UUR~(\ref{uur}) generates an overlap uncertainty relation for $n+1$ states, explored further in the Supplemental Material~\cite{sm}.

\paragraph{Experimental setup.---}
Our experiment uses polarisation states of single photons and a displaced Sagnac interferometer with controllable unitary transformations, $U$ in the  transmitted arm and $V$ in the reflected arm (see Fig.~\ref{fig:expt_setup}). 
We can determine the value of $\langle U^\dagger V\rangle$ for an input state $\rho$  by first noting that the average output photon number is given by
$\langle N\rangle_\chi = \half\left[ 1 + {\rm Re} \left\{e^{-i\chi} \langle U^\dagger V\rangle\right\}\right]$,
where $\chi$ is the phase difference between the two arms. Hence, we can obtain an interference pattern by varying $\chi$, with associated visibility
\beq \label{vis}
{\cal V}(U,V) := \frac{ \langle N\rangle_{max}- \langle N\rangle_{min}}{ \langle N\rangle_{max}+ \langle N\rangle_{min}} = |\langle U^\dagger V\rangle| . 
 \eeq
The values of $|\langle U\rangle|$ and $|\langle V\rangle|$ are similarly determined from the corresponding visibilities ${\cal V}(U,I), {\cal V}(I,V)$,
where $I$  denotes the identity transformation.  
Further, the phase of $\langle U^\dagger V\rangle$ corresponds to the value of the phase difference $\chi$ that gives maximum   average output photon number  (for a pure input state $|\psi\rangle$ this value is the Pancharatnam phase between $U|\psi\rangle$ and $V|\psi\rangle$~\cite{panch,loredo}). If $\chi(U,V)$ denotes the location of the interference maximum relative to some fixed phase reference value $\chi_0$, it follows that the phase of  $\langle U^\dagger V\rangle$ is given by
\beq \label{arg}
\arg \, \langle U^\dagger V\rangle = \chi(U,V) - \chi(I,I).
\eeq

Thus, our setup allows us to extract $\langle U^\dagger V\rangle$ from the interference pattern via Eqs.~(\ref{vis}) and~(\ref{arg}). More generally, this setup allows the Gram matrix coefficients $G_{jk}$ in Eq.~(\ref{uur}) to be experimentally determined for any set of unitary transformations $U_0=I,U_1,\dots U_n$ and polarisation state $\rho$, and hence the testing of the UUR for any $n$. We note that, in comparison, a recent qutrit experiment \blk testing a special case of \blk the $n=2$ UUR requires preparation of a strictly pure state $|\psi\rangle$, prior knowledge of the unitary operators (to implement both $V$ and $V^\dagger$), and tomographic reconstruction of $|\psi\rangle$, $U|\psi\rangle$ and $V|\psi\rangle$~\cite{express}.

As shown in Fig.~\ref{fig:expt_setup}, the main component of our setup is the displaced Sagnac interferometer, which is used to measure visibilities and phases as above.  For the single-photon source, we use a 410 nm  continuous-wave  diode laser to pump  an optically nonlinear beta  Barium Borate (BBO) crystal. The  degenerate  photon pairs generated by the non-collinear type-I spontaneous parametric down conversion  (SPDC)  are collected into optical fibres. The idler photon   heralds  the presence of  a signal photon. A half-waveplate (HWP) allows for a range of polarisation qubit states to be encoded on the signal photon, which is then sent  into the interferometer. 
  Each unitary operator, $U$ and $V$, is  implemented by a combination of  HWPs  and quarter-waveplates (QWP) , arranged in a group of four: 
HWP/QWP/HWP/QWP (Fig.~\ref{fig:expt_setup}),  with the QWPs set to 45$^{\circ}$ and the HWPs at variable angles $\alpha$ and $\beta$.  The condition  $\alpha=90\textdegree$ and $\beta=0\textdegree$  corresponds to  implementing the identity operation.  
 To realise an adjustable phase shift, a glass   element is  mounted on a motorised tilt controller and inserted in one arm of the interferometer. A fixed glass element is  positioned in the other arm in order to keep the path-length differences to within the  coherence length.  
Finally the photons are detected by silicon avalanche photodiodes.

\begin{figure}[t]
	\centering
	\includegraphics[keepaspectratio, width=8.65cm]{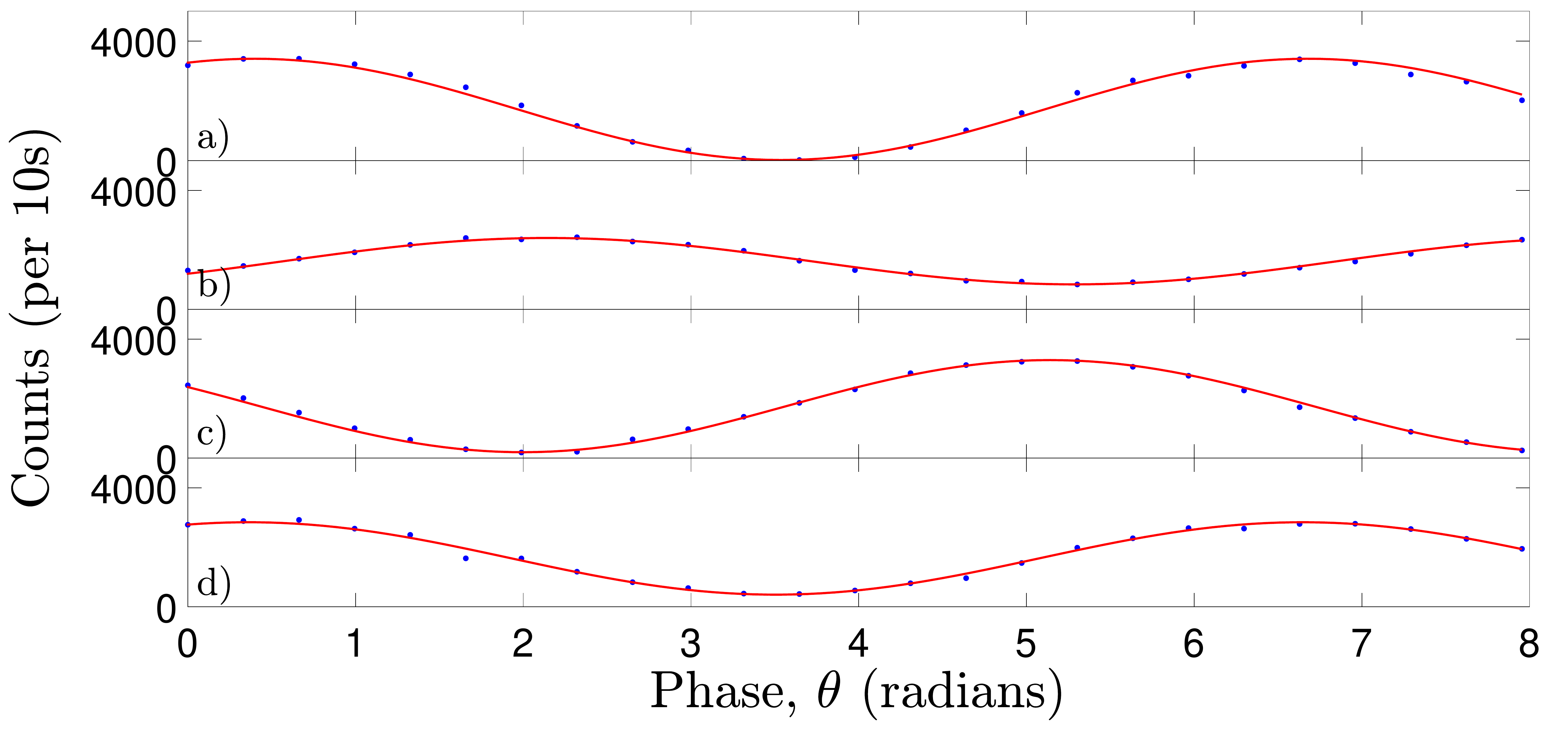}
	\caption{  Interferograms as recorded by measuring coincidence counts as a function of the applied phase shift.  Unitaries $U$, $V$ or $I$  are applied to each arm of the Sagnac interferometer  as follows  a) Transmitted: $I$, Reflected: $I$. b) Transmitted: $U$, Reflected: $I$. c) Transmitted: $I$, Reflected: $V$. d) Transmitted: $U$, Reflected:  $V$. The $|\langle U^\dagger V\rangle|$ terms or transition probabilities in the uncertainty relations can be calculated from the visibilities of the curves via Eq.~(\ref{vis}), and the Bargmann phase  via the phase terms in Eq.~(\ref{bargmann}), each of which is determined from the phase of a fringe pattern.}
	\label{fig:fitted_data}
\end{figure}

\paragraph{Results.---}
The interference fringes in Fig.~\ref{fig:fitted_data} are obtained by measuring the photon counts at the output of the interferometer.  The waveplates in one arm of the interferometer are rotated to produce either the identity operation $I$ or   an  operation  $U(\alpha_U,\beta_U)$   specified by $\alpha_U$ and $\beta_U$. Similarly, $V(\alpha_V,\beta_V)$ or $I$ can be implemented in the other arm. 
The left- and right-hand sides of the UUR in Eq.~(\ref{bargen}) are calculated by measuring four interference fringes, as shown in Fig.~\ref{fig:fitted_data}. We extract the phase and the visibility of the interference fringes by fitting the data to $A_1 + A_2\cos^2[\half(\theta-\theta_0)]$, where  $\theta=\chi-\chi_0$  is  the controlled phase shift implemented by the tilted glass   element.  
The visibility is then given by ${\cal V}(U,V) =A_2/(2A_1+A_2) $, and $\chi(U,V)$ by $\theta_0$.
The  phase $\Phi$ of the generalised Bargmann invariant $\langle U\rangle \langle U^\dagger V\rangle\langle V^\dagger\rangle$ follows via Eq.~(\ref{arg}) as 
\beq \label{bargmann}
\Phi = \chi(U,V) - \chi(U,I) - \chi(I,V)+ \chi(I,I).
\eeq

\begin{figure}[t]
	\centering
	\includegraphics[keepaspectratio, width= 8.65cm]{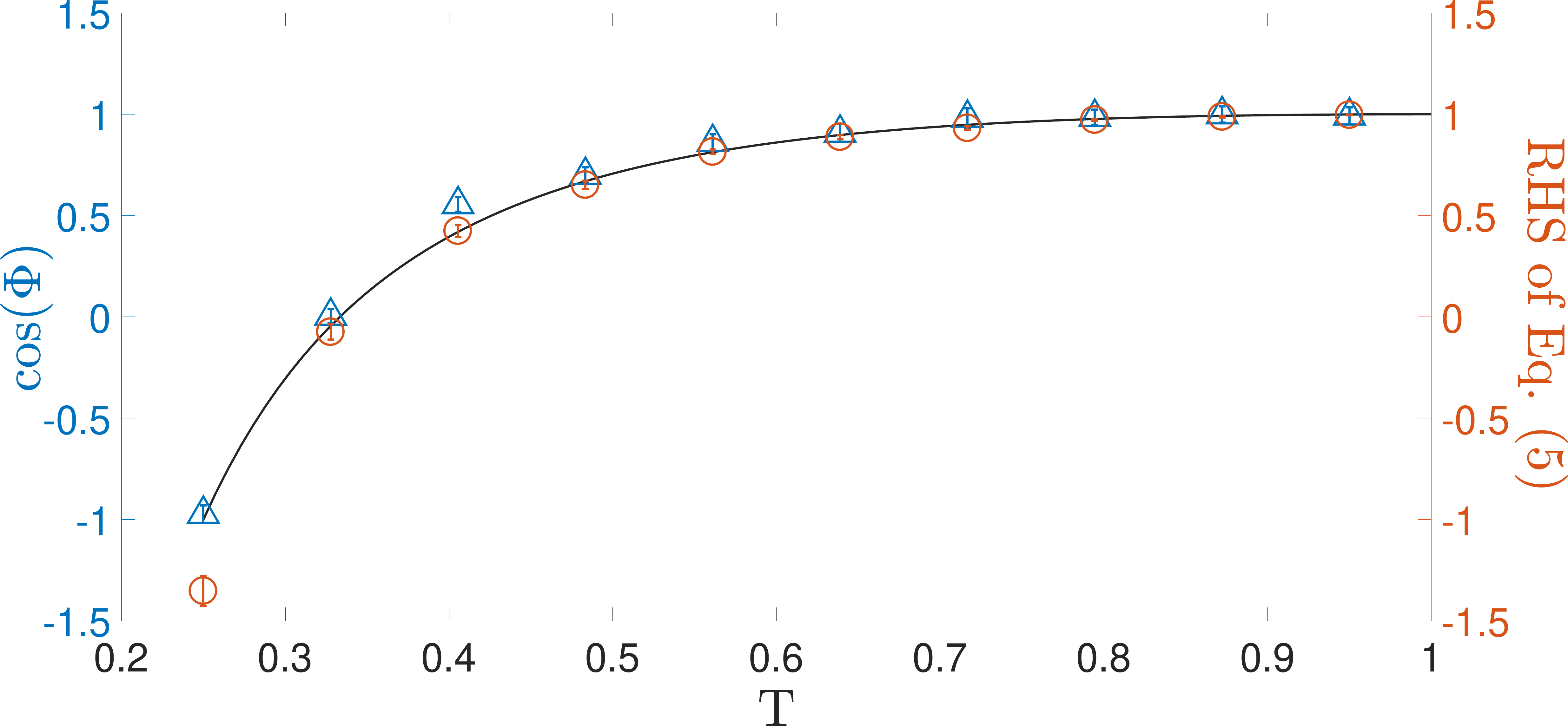}
	\caption{Verification of the UUR in the form of Eq.~(\ref{bargen}),  and its saturation by a pure qubit state.  
	 Using a fixed input state,   $U$ and $V$ are varied pairwise over a range as described in the text. At each step, parameterised by the mod-squared expectation value $T$ of the operators, the left-hand (blue triangles) and right-hand (red circles) sides of the UUR are determined from experimental measurements. The black line represents the theoretical prediction for saturation of Eq.~(\ref{bargen}).   The errors \blu are \blk calculated from the standard errors of the fit parameters. }
	\label{fig:bloch}
\end{figure}

Verification of the UUR in Eq.~(\ref{bargen}) can be seen in Fig.~\ref{fig:bloch}. Here, we fix the input state to be horizontally   polarised,   $\ket{H}$. We  vary $U$ and $V$  pairwise in steps such that the full range of $\cos\Phi$ is sampled and, at each setting, the tips of the Bloch vectors for $\{\ket{H},U\ket{H},V\ket{H}\}$ form an equilateral spherical triangle. This corresponds to $T_{12}=T_{23}=T_{13}$ in Eq.~(\ref{barg}).  Saturation of Eq.~(\ref{barg}) by pure qubit states corresponds to the area of the triangle being equal to $\Phi/2$~\cite{sm,mukunda,berry,aharonov}). 
In practice, there are small experimental imperfections. Although the states have high purity they are not completely mixture free, and so Eq.~(\ref{bargen}) replaces Eq.~(\ref{barg}) as the relevant UUR; also the nominal equilateral configuration is not exact. Nevertheless, since $|\langle U \rangle|^2 \approx |\langle V \rangle|^2 \approx |\langle U^\dag V \rangle|^2$, we write the average of these quantities as $T$, which forms the x-axis in Fig.~\ref{fig:bloch}. In the ideal pure state case for this configuration, $T=T_{ij}$ $(\forall i\neq j)$ and $T_{12}=|\langle U \rangle|^2$, etc.    

\begin{figure}[t!]
	\centering
	\includegraphics[keepaspectratio, width = 8.65cm]{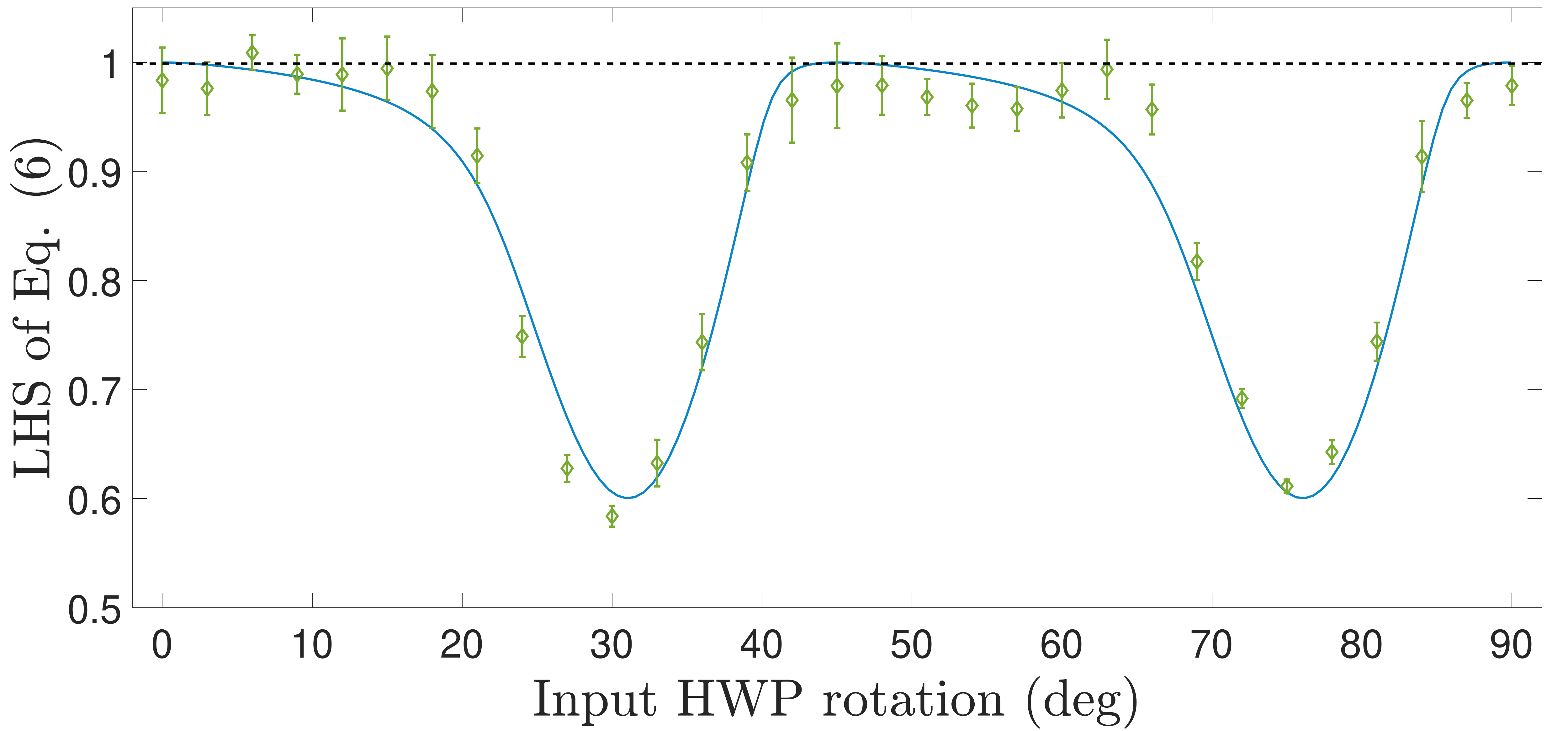}
	\caption{Experimental test of the OUR in Eq.~(\ref{our}). The unitary operators in both arms of the interferometer remain fixed and we input a family of states lying in the linear polarisation plane. The green data  points represent  the left-hand side of Eq.~(\ref{our}). The blue solid line represents the theoretical curve of the left-hand side of Eq.~(\ref{our}). Minimum uncertainty states correspond to a value of unity,  which is marked by the black dashed line. The error bars are calculated from the standard errors of the fit parameters. }
	\label{fig:mus}
\end{figure}

The experimental procedure to test the OUR in Eq.~(\ref{our}) uses a set of linearly polarised input states  of high purity,  and  fixed $U$($\alpha_U = 36^{\circ}, \beta_U = 0^{\circ}$) and $V$($\alpha_V = 0^{\circ}, \beta_V = 36^{\circ}$).
The transition probabilities in Eq.~(\ref{our})  may be determined from the measured visibilities in Eq.~(\ref{vis}), via $T_{jk}=|\langle U_j^\dagger U_k\rangle|^2={\cal V}(U_j,U_k)^2$ for a pure input state. The results are shown in Fig.~\ref{fig:mus}. The minimum uncertainty states correspond to the upper bound of unity in Eq.~(\ref{our}). 
 We note that one of the sources of error in our experiments is   the  imperfect calibration and retardation of the waveplates, which leads to the implemented unitary operations deviating slightly from the expected settings. 

\paragraph{Out-of-time-order correlators.---}  

  The UUR   may also be used to obtain a bound for the out-of-time-order correlator (OTOC), $F=\langle W_t^\dagger V^\dagger W_t V\rangle$, for a fixed unitary $V$ and time-dependent unitary $W_t$.  The OTOC determines the disturbance caused by $V$ on a later measurement of $W$ and, as noted in the Introduction, is of interest in quantum thermalisation, chaos and information scrambling, both in many  body and black hole physics~\cite{hosur, mald, swingle}.    It has   only very recently been experimentally measured for some systems~\cite{otocexp1,otocexp2,otocexp3}.

In particular, the UUR in Eq.~(\ref{uv}) implies 
$(1-u^2)(1-v^2)\geq(|\langle U^\dagger V\rangle|-uv)^2$, with $u=|\langle U\rangle|$, $v=|\langle V\rangle|$, yielding
\beq \label{ourgen}
|\langle U^\dagger V\rangle|\leq uv +\sqrt{(1-u^2)(1-v^2)} = \cos(\theta_U-\theta_V) ,
\eeq
with $\theta_U:=\cos^{-1} |\langle U\rangle|$.   Replacing $U$ by $W_tV$ and $V$ by $VW_t$ then gives the upper bound
\beq \label{oto1}
|F|=|\langle W_t^\dagger V^\dagger W_t V\rangle|\leq \cos(\theta_{VW_t} - \theta_{W_t V}) 
\eeq
for the modulus of the OTOC, which shows that $|F|$ is a direct signature of the noncommutativity of $V$ and $W_t$. Indeed, using   ${\rm Re} \{ F \} \leq |F|$   yields the lower bound
\beq \label{oto2}
\langle |[V,W_t]|^2\rangle = 2(1-{\rm Re}\, \{F\}) \geq 4\sin^2 \left(\frac{\theta_{VW_t} - \theta_{W_t V}}{2}\right) ,
\eeq
\blu where $\langle |A|^2\rangle$ denotes $\langle A^\dagger A\rangle$. \blk For polarisation qubits we note that the values of $\theta_{VW_t}$ and  $\theta_{W_t V}$ could be obtained from interferometer visibilities  corresponding to $|\langle VW_t\rangle|$ and $|\langle W_tV\rangle|$,  via Eq.~(\ref{vis}),  with a time-dependent unitary in one arm.

\paragraph{Conclusion.---}
We have presented a strong  and very general  unitary uncertainty relation (UUR), which implies the Robertson-Schr\"odinger relation and generates a tight  state overlap uncertainty relation. We tested these experimentally using polarisation qubit states in an interferometric configuration. This allowed for measurements that led directly to the quantities in the relation, and directly revealed the role of geometric phase in the UUR. We note that the UUR does not assume or require pure states, making it a general and powerful tool for real-world quantum systems.

We expect that the strength of the general UUR in Eq.~(\ref{uur}) will lead to further results that enhance and unify quantum uncertainty relations.   For example, noting that spin-1/2 observables are both Hermitian and unitary, the UUR in Eq.~(\ref{uv}) leads directly to, and hence encompasses, a tight state-independent qubit uncertainty relation obtained recently in Refs.~\cite{li,branciard}, and leads to a generalisation of the uncertainty relation for characteristic functions in Ref.~(\cite{rudnicki}) (see Supplemental Material~\cite{sm}).  It would also be of interest in future work to investigate possible connections of the UUR with \blu entropic, \blk measurement-disturbance and joint-measurement uncertainty relations; 
to test the UUR and OUR for higher values of $n$; and to implement similar tests of the OTOC bounds in Eqs.~(\ref{oto1}) and~(\ref{oto2}) above. 

\acknowledgments
We thank J. Kaniewski for pointing out the connection between the UUR and the uncertainty relation in Ref.~\cite{jed}. This work was supported by the ARC Centre of Excellence CE110001027.   S.W. acknowledges financial support through an Australian Government Research Training Program Scholarship.

%\newpage 
%~~
\newpage

%\onecolumngrid

\setcounter{equation}{0}
\renewcommand{\theequation}{S.\arabic{equation}}

\setcounter{figure}{0}
\renewcommand{\thefigure}{S\arabic{figure}}

\setcounter{page}{1}
\renewcommand{\thepage}{Supplemental Material -- \arabic{page}/5}

\section{\large{SUPPLEMENTAL MATERIAL}}

\section{I. Properties of the unitary uncertainty relation}

\subsection{A. An alternate proof}

We start by giving an alternate proof of the UUR in Eq.~(1) of the main text, using a matrix of operators. First, for a pure state $|\psi\rangle\langle\psi|$ and unitaries $U_0=I, U_1,U_2,\dots,U_n$, define the kets $|\psi_j\rangle:=U_j|\psi\rangle$.  Defining the Gram matrix $G(\psi)$ via the coefficients $G_{jk}(\psi):=\langle\psi_j|\psi_k\rangle$, it immediately follows that
\beq \label{gpsi}
G(\psi) \geq 0 .
\eeq
This already implies the UUR for the case of pure states. Note that it can also be written in matrix operator form as
\beq \label{hatg}
\hat G \geq 0,\qquad   \hat G_{jk} := U_j^\dagger U_k .
\eeq
It follows from either of the above two equations that 
\beq \label{grho}
G(\rho)\geq 0, \qquad G(\rho):=\sum_k p_k G(\psi_k)=\tr{\rho \hat G},
\eeq 
for any density operator $\rho=\sum_k p_k|\psi_k\rangle\langle\psi_k|$, yielding the general UUR
\beq \label{uursm}
\det G(\rho) \geq 0
\eeq
as desired. 

  Note that it is straightforward to formally generalise the UUR  to arbitrary (not necessarily Hermitian) operators $A_1, A_2,\dots A_n$, with $v^{(j)}$ of the main text and $|\psi_j\rangle$ above replaced by $A_j\rho^{1/2}$ and $A_j|\psi\rangle$, respectively.  

\subsection{B. Deriving the Robertson-Schr\"odinger relation}

For $n=2$, writing $U_1=U$ and $U_2=V$, (\ref{uursm}) reduces to
\beq \label{det}
\left| \begin{array}{ccc}
	1&\langle U\rangle &\langle V\rangle\\
	\langle U^\dagger\rangle & 1 &\langle U^\dagger V\rangle\\
	\langle V^\dagger\rangle & \langle V^\dagger U\rangle & 1
\end{array} \right|
\geq 0, 
\eeq
which simplifies to
\beq \label{uvsm}
{\rm Var}\,U\,{\rm Var}\,V \geq |\langle U^\dagger V\rangle - \langle U^\dagger\rangle\langle V\rangle|^2 
\eeq
as per Eq.~(2) of the main text. Choosing $U=e^{i\epsilon A}$ and $V=e^{i\epsilon B}$ for any two Hermitian operators $A$ and $B$, one has the Taylor expansions
\[ U=I+i\epsilon A-\half \epsilon^2 A^2+O(\epsilon^3),~~V=I+i\epsilon B-\half \epsilon^2 B^2+O(\epsilon^3),\]
from which it follows that
\begin{align}
{\rm Var}\, U&:=1-|\langle U\rangle|^2 = \epsilon^2\, {\rm Var} \,A +O(\epsilon^3),\\
{\rm Var}\, V&:=1-|\langle V\rangle|^2 = \epsilon^2\, {\rm Var} \,B +O(\epsilon^3),
\end{align}
and
\begin{align}
|\langle U^\dagger V\rangle - \langle U^\dagger\rangle\langle V\rangle|^2 &= \epsilon^4\left|\langle AB\rangle-\langle A\rangle\langle B\rangle\right|^2 + O(\epsilon^5)\nn\\ 
&= \epsilon^4\big|{\rm Re}\{\langle AB\rangle\}-\langle A\rangle\langle B\rangle\nn\\
&\qquad~~ + i\,{\rm Im}\{\langle AB\rangle\}\big|^2 + O(\epsilon^5)\nn\\
&= \blu \epsilon^4 \left[ \left( \langle \half(AB+BA)\rangle -\langle A\rangle\langle B\rangle\right)^2 \right. \blk \nn\\
&\qquad~~ \blu  +\left. \langle \half(AB-BA)\rangle^2\right] + O(\epsilon^5) \blk 
\nn\\
&= \epsilon^4\left( {\rm Cov}(A,B)^2 + \frac{1}{4}  |\langle [A,B]\rangle|^2\right) \nn\\
&\qquad\qquad +O(\epsilon^5) .
\end{align}
Substituting these results into Eq.~(\ref{uvsm}), dividing by $\epsilon^4$, and taking the limit $\epsilon\rightarrow0$, then yields the Robertson-Schr\"odinger uncertainty relation
\beq  \label{rssm}
{\rm Var}A\,{\rm Var}B \geq \frac{1}{4}|\langle [ A, B]\rangle|^2 + {\rm Cov}(A,B)^2,
\eeq
as per Eq.~(3) of the main text.  Note that one cannot proceed in the reverse direction, i.e., the UUR is strictly stronger than the Robertson-Schr\"odinger uncertainty relation.    As shown further below, the UUR may also be used to derive a tight state-independent qubit uncertainty relation~\cite{li,branciard}.

\subsection{C. Minimum uncertainty qubit states}

It was shown in the main text that the UUR is saturated by all pure states of any $d$-dimensional Hilbert space with $d\leq n$.  \blu This is a sign of the strength of this uncertainty relation.  For example, the standard Heisenberg uncertainty relation for position and momentum is only saturated by a particular subset of Gaussian pure states, whereas the stronger Robertson-Schr\"odinger uncertainty relation is saturated by {\it all} Gaussian pure states. It is precisely the strength of our relation that enables us to derive many other uncertainty relations in the literature as corollaries. \blk 

More generally, as noted in the main text, the UUR in  Eq.~(1) of the main text (repeated in Eq.~(\ref{uursm}) above)  is saturated if and only if the $n+1$ operators $v^{(j)}=U_j\rho^{1/2}$ are linearly \blu dependent. \blk Since there are at most $d^2$ linearly independent operators in a $d$-dimensional Hilbert space, this implies in particular that the UUR is saturated by all density operators when $d^2\leq n$.  For qubits, i.e., $d=2$, this raises the interesting question of whether there are any density operators, apart from pure states, that saturate the UUR for $n=2$ or $n=3$. 

It turns out that this question has a simple answer when $n=2$: a non-pure qubit density operator $\rho$ saturates the UUR in Eq.~(\ref{uvsm}) if and only if $[U,V]=0$, i.e., if and only if $U$ and $V$ correspond to rotations about the same axis of the Bloch sphere. Hence, {\it all} qubit states are  minimum uncertainty states of Eq.~(\ref{uvsm}) if $U$ and $V$  commute, while only all {\it pure} qubit states are minimum uncertainty states if they do not.

To prove this result, first note that if $\rho$ is a non-pure qubit state then it is invertible, so that $v^{(j)}$ may be multiplied on the right by $\rho^{-1/2}$.  Hence, saturation of Eq.~(\ref{uvsm}) is equivalent to linear \blu dependence \blk of the unitary operators $I, U, V$, i.e., to
\beq \label{lindep}
\alpha I+\beta U +\gamma V=0
\eeq
for some constants $(\alpha,\beta,\gamma)\neq(0,0,0)$. Taking the commutator with $U$ or $V$ leads to $\beta[U,V]=0=\gamma[U,V]$. But if $\beta$ and $\gamma$ both vanish then the above equation yields $\alpha=0$. Hence linear independence implies $[U,V]=0$. Conversely, if $U$ and $V$ commute we have
\beq 
U=e^{i\chi_0}|0\rangle\langle0 | + e^{i\chi_1}|1\rangle\langle 1|,~V=e^{i\kappa_0}|0\rangle\langle0 | + e^{i\kappa_1}|1\rangle\langle 1|
\eeq 
for some qubit basis $\{|0\rangle,|1\rangle\}$ and phases $\chi_0,\chi_1,\kappa_0,\kappa_1$. Substituting into Eq.~(\ref{lindep}) then gives
\beq
\left( \begin{array}{cc}
	e^{i\chi_0} & e^{i\kappa_0} \\
	e^{i\chi_1} & e^{i\kappa_1}
\end{array} \right)
\left( \begin{array}{c}
	\beta \\
	\gamma
		\end{array} \right)
= -\left( \begin{array}{c}
	\alpha \\
	\alpha
		\end{array} \right) .
\eeq
This clearly has nontrivial solutions  $(\alpha,\beta,\gamma)\neq(0,0,0)$, corresponding to linear dependence, when the determinant of the matrix on the left-hand side does not vanish, i.e., if $e^{i(\chi_1-\chi_0)}\neq e^{i(\kappa_1-\kappa_0)}$. Finally, if the determinant does vanish, then one still has linear dependence since $U=e^{i\phi}V$ in this case (with $\phi=\chi_0-\kappa_0$).  

\blu The above result generalises straightforwardly to higher dimensions and numbers of unitary operators whenever $\rho$ is invertible. That is, an invertible density operator $\rho$ saturates the UUR for $n$ unitary operators if and only if $[U_j,U_k]=0$ for $j,k=1,2,\dots,n$. In particular, saturation of the UUR in Eq.~(\ref{uursm}), $\det G(\rho)=0$, is then equivalent to $\alpha_0 I+\sum_j {\alpha_j} U_j=0$, generalising Eq.~(\ref{lindep}). \blk
  
\subsection{D. Tight state-independent qubit uncertainty relation}

Qubit observables may be represented, up to translation and rescaling, by Hermitian operators of the form $A=\bm a\cdot\sigma$, where $\bm a$ is a unit 3-vector and $\sigma\equiv(\sigma_1,\sigma_2,\sigma_3)$ denotes the vector of Pauli operators. It follows that $A^\dagger A=A^2=I$, i.e., these operators are unitary.  Hence the UUR may be directly applied to such observables. Further, noting the above section, the corresponding uncertainty relations are saturated for all pure qubit states.

For example, for $n=2$ consider the UUR in Eq.~(2) of the main text, i.e., as per Eqs.~(\ref{det}) and (\ref{uvsm}), for the case of a qubit state $\rho=\half(I+\bm r\cdot \sigma)$. Replacing $U$ and $V$ by $A=\bm a\cdot\sigma$ and $B=\bm b\cdot\sigma$, respectively, for arbitrary unit directions $\bm a, \bm b$, and using
\beq \label{qubituv}
 \langle U^\dagger V\rangle = \langle AB\rangle = \bm a\cdot \bm b + i(\bm a\times\bm b)\cdot \bm r ,
 \eeq
leads via Eq.~(\ref{det}) to
\beq \label{abid}
\langle A\rangle^2+ \langle B\rangle^2 -2(\bm a\cdot\bm b)\langle A\rangle\langle B\rangle \leq 1-(\bm a\cdot\bm b)^2-[(\bm a\times\bm b)\cdot \bm r]^2  ,
\eeq
with equality for pure states, i.e., when $|\bm r|=1$.  It follows that one has the tight state-independent uncertainty relation
\begin{align}
(\Delta A)^2 +& (\Delta B)^2 + 2|\bm a\cdot\bm b|\sqrt{1-(\Delta A)^2}\sqrt{1-(\Delta B)^2} \nn\\
&\geq 2 -\left[\langle A\rangle^2 + \langle B\rangle^2 -2(\bm a\cdot\bm b)\langle A\rangle\langle B\rangle\right] \nn\\
&\geq 1+(\bm a\cdot\bm b)^2+[(\bm a\times\bm b)\cdot \bm r]^2 \nn \\
&\geq 1+(\bm a\cdot\bm b)^2  \label{cyril}
\end{align}
\blu for $\Delta A$ and $\Delta B$, \blk given for particular cases in~\cite{li} and more generally in~\cite{branciard}. \blu Here the second line follows using $(\Delta A)^2=1-\langle A\rangle^2$ and $|x|\geq -x$, and the third line via Eq.~\ref{abid}).
\blk The $n=3$ case is considered in Sec.~I.F below. \blk
 
 \blk
\subsection{E. Connection to other uncertainty relations}

The uncertainty relation for characteristic functions of the position and momentum observables $X$ and $P$ in Eq.~(10) of Ref.~\cite{rudnicki}, used to derive theorem~1 therein, is a special case of the UUR for $n=2$, corresponding to the choice $U\equiv e^{i\lambda_x X}$, $V=e^{i\lambda_p P}$. Moreover, this may be generalised, via the UUR for arbitrary $n$, to  an uncertainty relation for the characteristic function $\chi(\theta,\tau)=\langle e^{i(\theta X+\tau P)/\hbar}\rangle$ of the Wigner quasiprobability function, via the choice $U_j=e^{i(\theta_j X+\tau_j P)/\hbar}$, $j=1,2,\dots,n$. 
	
The positivity of the matrix $G(\rho)$ in Eq.~(\ref{grho}) for a given $n$ is equivalent to positivity of the determinants of its principal minors~\cite{gram}, i.e., to the UURs for $1, 2,\dots,n$ unitary operators. By Schur's lemma, it is also equivalent to the $n\times n$ matrix inequality~\cite{gram}
\beq \label{uuc}
\bm u \bm u^\dagger \leq C, 
\eeq
where the vector $\bm u$ and correlation matrix $C$ are defined by $u_j:=\langle U_j\rangle$, $C_{jk}:=\langle U^\dagger_jU_k\rangle$, $j,k=1,2,\dots,n$. This may be regarded as a vector form of the Schwarz inequality, and becomes formally equivalent to Robertson's uncertainty relation for $n$ Hermitian operators if the $U_j$ are replaced by Hermitian operators $A_j$~\cite{rob34} (see also the last paragraph of Sec.~I.A above). With this replacement it is also a generalisation of the ellipsoid condition in Ref.~\cite{jed}, used there to obtain strong uncertainty relations for anticommuting Hermitian observables. For qubit observables, as in the previous section, which are both unitary and Hermitian, there is thus a direct correspondence.  Note that for such observables the ellipsoid condition implies the lemmas used in Ref.~\cite{branciard} to obtain state-independent qubit uncertainty relations, such as Eq.~(\ref{cyril}) in Sec.~II.D above.

\blk
\subsection{F. Physical invariance and Bargmann invariants}

A unitary transformation $U$ takes any given state $\rho$ to the state $U\rho U^\dagger$. Hence, $U$ and $e^{i\phi}U$ are physically equivalent transformations for any phase $\phi$. As remarked in the main text, the general UUR in Eq.~(\ref{uursm}) respects this physical equivalence.  In particular, under $U_j\rightarrow e^{i\phi_j}U_j$, it follows from Eqs.~(\ref{hatg}) and~(\ref{grho}) that
\beq
G(\rho) \rightarrow K G(\rho) K^\dagger
\eeq
where $K$ is the unitary diagonal matrix
\beq
K:=\left( \begin{array}{cccc}
	e^{-i\phi_0} & 0 & \dots & 0\\
		0 & e^{-i\phi_1} & \dots & 0\\
		\vdots & \vdots & \ddots & \vdots\\
		0 & 0 & \dots & e^{-i\phi_n}
\end{array} \right) .
\eeq
Hence the positivity of $G(\rho)$ is preserved (as are its eigenvalues), implying that the UUR respects physical invariance as claimed in the main text.

In fact, the UUR satisfies a stronger form of physical invariance. In particular, expanding the determinant in Eq.~(\ref{uursm}) in the usual way, the UUR can be rewritten as a sum of products,
\beq \label{perm}
\sum_P (-1)^P \langle U_0 U^\dagger_{P(0)} \rangle\dots \langle U_nU^\dagger_{P(n)}\rangle \geq 0,
		\eeq
where the sum is over permutations of $\{0,1,...,n\}$ and $(-1)^P$ denotes the sign of permutation $P$. In each summand, note that $U_j$ and $U_j^\dagger$ will appear exactly once, for each value of $j$.  Hence, each product is invariant under $U_j\rightarrow e^{i\phi_j}U_j$, i.e., each individual term in the above sum respects physical invariance.

This stronger property is closely related to Bargmann invariants~\cite{barg}.  The Bargmann invariant associated with a given set of $m$ pure states $|\psi_1\rangle,\dots |\psi_m\rangle$ is defined by the cyclic product 
$B_{12\dots m}:= \langle \psi_1|\psi_2\rangle \langle \psi_2|\psi_3\rangle\dots \langle \psi_m|\psi_1\rangle$, which is clearly invariant under rephasings $|\psi_j\rangle\rightarrow e^{i\phi}|\psi_j\rangle$. Thus it is a projective invariant, as may also be seen by writing it as a trace of a product of projection operators, $B_{1\dots m}=\tr{|\psi_1\rangle\langle \psi_1|~|\psi_2\rangle\langle \psi_2| \dots|\psi_m\rangle\langle \psi_m|}$.  Choosing $|\psi_{j+1}\rangle:=U_j|\psi\rangle$, it follows immediately that the UUR (\ref{perm}) can be rewritten as as uncertainty relation for the associated Bargmann invariants.

For example, for $n=2$ and pure state $\rho=|\psi\rangle\langle\psi|$ the UUR in Eq.~(\ref{det}) simplifies to
\beq \label{barg2}
1-B_{23}-B_{12}-B_{13}+B_{123}+B_{321} \geq 0,
\eeq
which is equivalent to Eq.~(4) of the main text. Similarly, for $n=3$ one finds via Eq.~(\ref{perm}) that the Bargmann invariants associated with any 4 pure states satisfy the uncertainty relation
\begin{align} \label{barg3}
2&\leq (1-B_{12})(1-B_{34}) + (1-B_{13})(1-B_{24})  \nn \\
&\qquad +(1-B_{14})(1-B_{23}) \nn\\
&\qquad +2\, {\rm Re}\left\{B_{123} + B_{124}+ B_{134} +B_{234}\right\} \nn\\
&\qquad - 2\, {\rm Re}\left\{B_{1234} + B_{1243} + B_{1324} \right\}.
\end{align}
 This relation may equivalently be written as
\begin{align} \label{barg4}
{\rm Var}\, U\,& {\rm Var} (V^\dagger W) + {\rm Var}\, V \,{\rm Var} (U^\dagger W) + {\rm Var} \,W\, {\rm Var} (U^\dagger V) \nn \\
&\geq 2 - 2\, {\rm Re}\left\{B_{123} + B_{124}+ B_{134} +B_{234}\right\} \nn\\
&~~~~~~~+ 2\, {\rm Re}\left\{B_{1234} + B_{1243} + B_{1324} \right\},
\end{align}
in terms of the variances of $U$, $V$ and $W$,   with equality holding for all qubit pure states as per the main text.  

More generally, for mixed states the UURs for $n=2$ and $n=3$ take exactly the same form as Eqs.~(\ref{barg2})-(\ref{barg4}), providing one defines the {\it generalised} Bargmann invariant by the cyclic product
\beq \label{barin}
B_{12\dots m}:= \langle U_1 U_2^\dagger\rangle \langle U_2 U_3^\dagger\rangle\dots\langle U_mU_1^\dagger\rangle
\eeq
for any given set of unitary operators $U_1,U_2,\dots,U_m$ and state $\rho$.
Note that $B_{12\dots m}$ may be measured for polarisation qubits via the interferometric method of the main text.   

\blk
Note for the qubit case that the left-hand side of Eq.~(\ref{barg4}) simplifies, using Eq.~(\ref{qubituv}), to
\beq
|\bm b\times \bm c|^2 (\Delta A)^2 +|\bm a\times \bm c|^2 (\Delta B)^2 +|\bm a\times \bm b|^2 (\Delta C)^2 ,
\eeq
for three qubit observables $A=\sigma\cdot \bm a$, $B=\sigma\cdot \bm b$, $C=\sigma\cdot \bm c$. This is the same as the first three terms of the state-independent uncertainty relation in Eq.~(35) of Ref.~\cite{branciard}, and indeed the latter can be derived via the equivalent relation~(\ref{uuc}) for such observables (also equivalent to the ellipsoid condition in Ref.~\cite{jed} for such observables).
\blk

\subsection{G. Spherical trigonometry and the geometric phase}

The saturation of the UUR in Eq.~(4) of the main text by all pure qubit states corresponds to the statement that
\beq \label{tri1}
\cos \Phi = \frac{T_{12}+T_{13}+T_{23} -1}{2\sqrt{T_{12}T_{13}T_{23}}}
\eeq
for the transition probabilities $T_{jk}=|\langle\psi_j|\psi_k\rangle|^2$ of any three qubit states $|\psi_1\rangle$, $|\psi_2\rangle$, $|\psi_3\rangle$, where $\Phi$ is the phase of the Bargmann invariant $B_{123}=\langle\psi_1|\psi_2\rangle\langle\psi_2|\psi_3\rangle\langle\psi_3|\psi_1\rangle$. 

\begin{figure}[t!]
	\centering
	\includegraphics[keepaspectratio,width = 8.65cm]{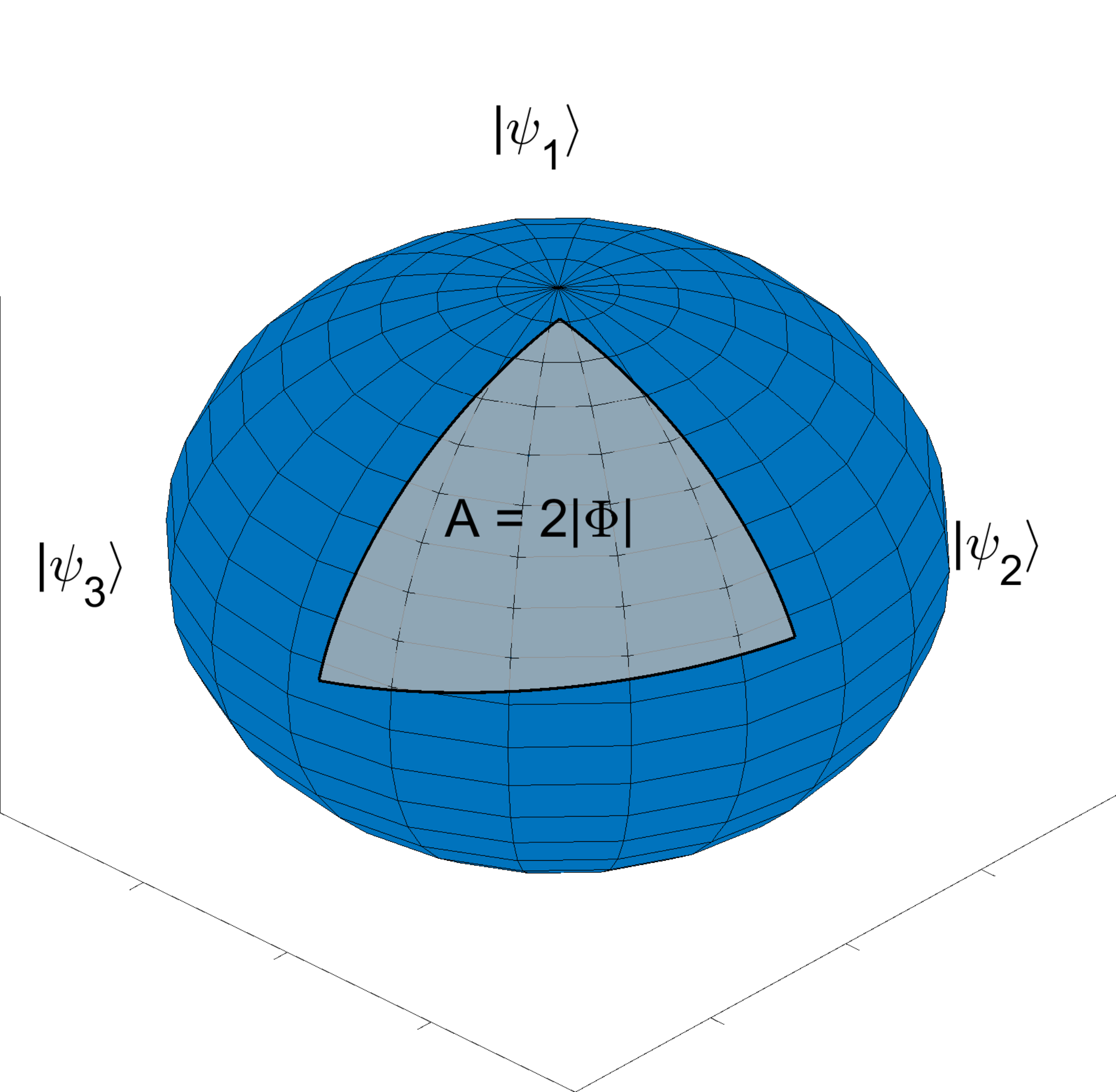}
	\caption{Equality in Eq.(\ref{tri1}, for any pure qubit states $|\psi_1\rangle, |\psi_2\rangle,|\psi_3\rangle$, corresponds to a simple geometric relation connecting the Bargmann phase $\Phi$ of the states to the area $A$ of the corresponding triangle on the Bloch sphere, as per Eq.~(\ref{area}).}
	\label{fig:triangle}
\end{figure}

Denoting the angles between the corresponding pairs of Bloch vectors by $\gamma_{jk}$ then $T_{jk}=\cos^2 \half\gamma_{jk}$, and the above equality can be rewritten as
\beq \label{tri2}
\cos\Phi =
 \frac{\cos^2{\half\gamma_{12}} + \cos^2{\half\gamma_{13}} + \cos^2{\half\gamma_{23}}-1}{2 \cos{\half\gamma_{12}}\cos{\half\gamma_{13}}\cos{\half\gamma_{23}}} =\cos \half A,
\eeq
where $A$ is the area of the spherical triangle formed by the Bloch vectors, and the second equality is an old but little known identity from spherical trigonometry (see Eq.~(3) of section~103 of~\cite{trig}).  Thus, saturation of the UUR by qubit states corresponds to the known relation~\cite{mukunda}
\beq \label{area}
A=2|\Phi|,
\eeq
between Bargmann phase and the area of a spherical triangle, depicted in Fig.~\ref{fig:triangle}. This relation is an example of the more general connection between area and geometric phase~\cite{mukunda}, where evolution of a qubit about a circuit on the Bloch sphere generates a geometric phase equal to half the area enclosed by the circuit~\cite{berry, aharonov}.

\section{II. Properties of the overlap uncertainty relation}

\subsection{A. Saturation}
\label{sat}

The overlap uncertainty relation in Eq.~(6) of the main text, 
\beq \label{ourcopy}
T_{12}+T_{13}+T_{23} - 2 \sqrt{T_{12}T_{13}T_{23}} \leq 1 ,
\eeq
 for the transition probabilities of any three pure states $|\psi_1\rangle, |\psi_2\rangle, |\psi_3\rangle$, may be rewritten as
\begin{align} \label{oursm}
1&\geq \cos^2{\theta_{12}} + \cos^2{\theta_{13}} +\cos^2{\theta_{23}} \nn\\
		&\qquad- 2 \cos{\theta_{12}}\cos{\theta_{13}}\cos{\theta_{23}} ,
\end{align}
where $\theta_{jk}=\cos^{-1} |\langle\psi_j|\psi_k\rangle| \in[0,\pi/2]$ is the angle between states $|\psi_j\rangle$ and $|\psi_k\rangle$. Saturation of this relation corresponds to a quadratic equation for $\cos \theta_{23}$, which is easily solved to give
\beq
\cos\theta_{23} = \cos(\theta_{12}\pm\theta_{13}).
\eeq
It follows, recalling $\theta_{jk}\in[0,\pi/2]$, that
\beq \label{theta}
\theta_{23} =|\theta_{12} \pm \theta_{13}| ,
\eeq
i.e., one of $\theta_{23} =\theta_{12} + \theta_{13}$, $\theta_{12} =\theta_{13} + \theta_{23}$, $\theta_{13} =\theta_{12} + \theta_{23}$ must hold.  Since the angle between two pure states is a metric (the well-known Fubini-Study metric), it follows from the triangle inequality that saturation corresponds to the states lying on a common geodesic with respect to this metric, as claimed in the main text. 

For the case of qubit states, one has the simple relation $\theta_{jk}=\half\gamma_{jk}$, where $\gamma_{jk}\in[0,\pi]$ is the angle between the corresponding Bloch vectors. Hence, in this case, Eq.~(\ref{theta}) immediately implies that saturation is equivalent to the Bloch vectors lying on a great semicircle of the Bloch sphere (corresponding to a spherical triangle of area  $A=0$).

\blk Note that Eq.~(\ref{theta}) implies that the possible values of $(\theta_{12}, \theta_{23}, \theta_{13})\in[0,\pi/2]^3$ are restricted to the polytope with vertices $(0,0,0)$, $(0,\pi/2,\pi/2)$, $(\pi/2,0,\pi/2)$, $(\pi/2,\pi/2,0)$ and $(\pi/2,\pi/2,\pi/2)$. As noted in the main text, this is a stronger restriction than inequality obtained from the triangle inequality for the trace distance between the states.  The latter inequality and its permutations may be rewritten  as
\beq \label{tracedist}
\sin\theta_{23} \leq |\sin\theta_{12} \pm \sin\theta_{13}|,
\eeq 
and hence allows values of $(\theta_{12}, \theta_{23}, \theta_{13})$ falling outside the above polytope (cf. Eq.~(\ref{theta})). For example, Eq.~(\ref{tracedist})  is satisfied by  the triple $(\pi/2,\pi/6,\pi/6)$. In contrast, this triple, corresponding to $T_{12}=0$, $T_{23}=T_{13}=3/4$, is excluded by the OUR in Eq.~(\ref{ourcopy}). \blk

\subsection{B. Minimum uncertainty states for fixed $U$ and $V$}

\blu Quantum uncertainty relations may be broadly viewed as constraints imposed by quantum mechanics on what is possible.  Classical mechanics places no such fundamental constraints, and more general theories, e.g., generalised probabilistic theories, typically place much weaker constraints.  In this sense we view the overlap uncertainty relations of the paper as being of physical interest because it captures a fundamental constraint on transition probabilities, where the latter are of course very important quantities in quantum theory.  \blk

For pure states $|\psi_1\rangle=|\psi\rangle, |\psi_2\rangle=U|\psi\rangle, |\psi_3\rangle=V|\psi\rangle$, the corresponding overlap uncertainty relation can be rewritten as the uncertainty relation
\beq
{\rm Var}\,U\,{\rm Var}\,V \geq \left( |\langle U^\dagger V\rangle| - |\langle U\rangle|\, |\langle V\rangle|\right)^2 
\eeq
for the unitary operators $U$ and $V$.  
The minimum uncertainty states corresponding to the saturation of this relation are, therefore, those states $|\psi\rangle$ for which $|\psi\rangle, U|\psi\rangle, V|\psi\rangle$ lie on a common geodesic in Hilbert space (see Sec.~II~A above). \blu This captures the `classical' property that $U$ and $V$ are effectively `commuting' for state $|\psi\rangle$, with
$U|\psi\rangle=e^{i s A}$ and $V|\psi\rangle =e^{itA}|\psi\rangle$ for some fixed $A$. \blk

From Eq.~(\ref{theta}),  the minimum uncertainty states are given by the solutions of
\beq \label{cos}
\cos^{-1} |\langle\psi|U^\dagger V|\psi\rangle| = \left| \cos^{-1} |\langle\psi|U|\psi\rangle| \pm\cos^{-1} |\langle\psi|V|\psi\rangle|   \right|. 
\eeq

\begin{figure}[t!]
	\centering
	\includegraphics[keepaspectratio, width=8.65cm]{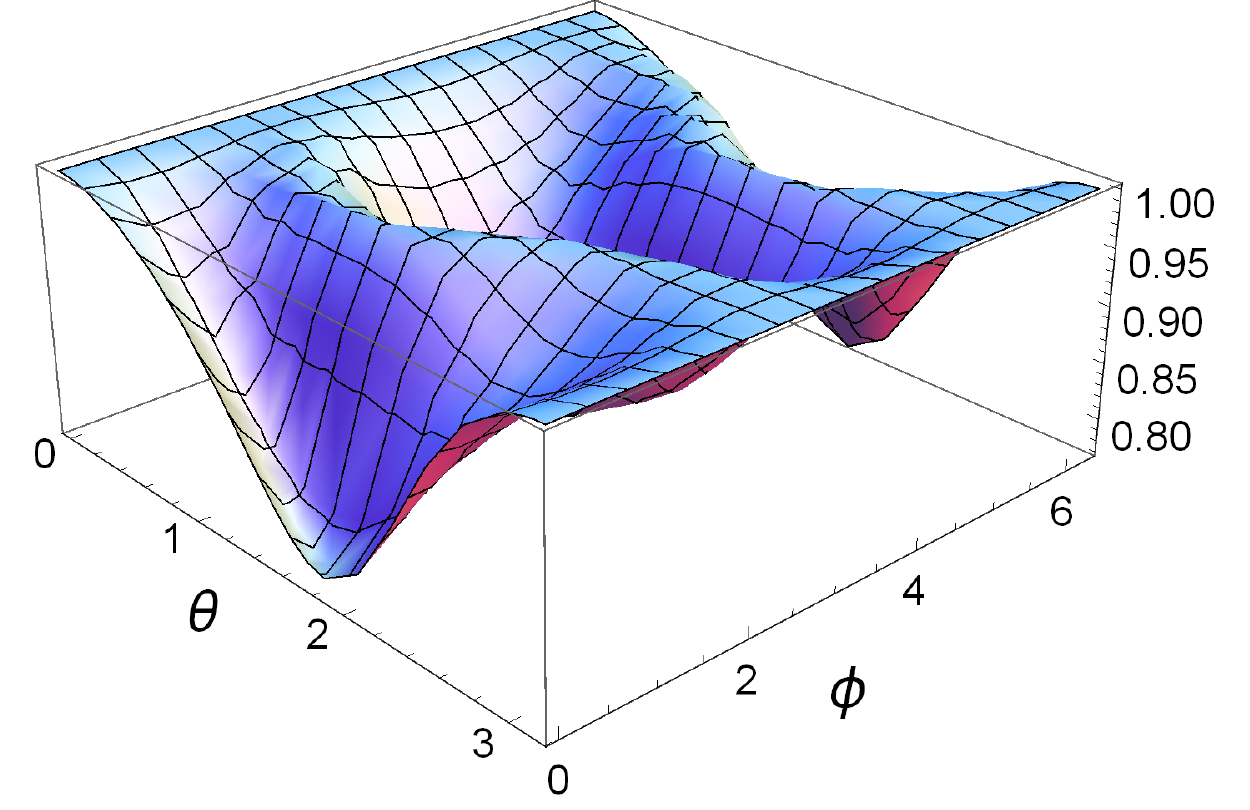}
	\caption{The OUR in Eq.~(\ref{ourcopy}) for fixed $U$ and $V$. The Bloch vector of state $|\psi\rangle$ is parameterised by spherical coordinates $(\theta,\phi)$, and the left-hand side of Eq.~(\ref{ourcopy}) is shown on the vertical axis. Saturation occurs for minimum uncertainty states, and it is seen from the figure that there are two families of such states, where these correspond to the cases that $|\psi\rangle, U|\psi\rangle$ and $V|\psi\rangle$ lie on a common great semicircle of the Bloch sphere. The choices for $U$ and $V$ in the figure are given by $U=e^{i\pi\sigma_y/8}$ and $V=e^{i\pi \sigma_z/8}$, corresponding to Bloch sphere rotations of $\pi/4$ about the $y$-axis and $z$-axis respectively.  The minimum uncertainty states correspond to a maximum value of unity. Note that Fig.~4 of the main text corresponds to a vertical cross section of a similar surface. }
	\label{fig:oursm}
\end{figure}

For qubit minimum uncertainty states, the Bloch vectors of $|\psi\rangle, U|\psi\rangle, V|\psi\rangle$  lie on a great semicircle of the Bloch sphere (see above).  Further, note that the angle $\gamma$ between two unit Bloch vectors $\bm a$ and $\bm b$ satisfies $\cos\gamma=\bm a\cdot \bm b$. Hence,   denoting the Bloch vector of $|\psi\rangle$ by  $\bm a$  and $U$ and $V$ by suitable rotation matrices $R_U$ and $R_V$, Eq.~(\ref{cos}) reduces to (recalling $\theta_{jk}=\half\gamma_{jk}$):
\beq
\cos^{-1}\frac{\bm a^\top R_U^\top R_V\bm a}{2} =\left| \cos^{-1}\frac{\bm a^\top R_U\bm a}{2}\pm \cos^{-1}\frac{\bm a^\top R_V\bm a}{2}\right| .
\eeq
Note that if $R_U$ and $R_V$ correspond to rotations about the unit vectors $\bm m$ and $\bm n$, then $a=\pm m$ and $\bm =\pm n$ are solutions of these equations.  More generally, since $\bm a$ has two continuous degrees of freedom, it follows that there will be in general two 1-parameter families of minimum uncertainty states, corresponding to the $+$ and $-$ signs respectively (although more solutions are possible in degenerate cases). A generic example is shown in Fig.~\ref{fig:oursm}.

\subsection{C. Higher-order OURs}

The OUR in Eq.~(6) of the main text (see also Eq.~(\ref{oursm}) above) follows from the UUR with $n=2$. One can similarly obtain OURs corresponding to larger values of $n$.  For example, noting that the definition of the Bargmann invariants in Eq~(\ref{barin}) \blu implies  $B_{j_1j_2\dots j_m}=\sqrt{T_{j_1j_2}\dots T_{j_{m-1}j_m}T_{j_mj_1}}$ \blk for pure states, \blk and using ${\rm Re}\{x+y\}\leq |x|+|y|$, the UUR for $n=3$ in Eq.~(\ref{barg3}) yields the OUR
\begin{align} \label{our3}
1&\leq \half(1-T_{12})(1-T_{34}) + \half(1-T_{13})(1-T_{24})  \nn \\
&\qquad +\half(1-T_{14})(1-T_{23}) \nn\\
&\qquad +\sqrt{T_{12}T_{23}T_{13}} + \sqrt{T_{12}T_{24}T_{14}}+ \sqrt{T_{13}T_{34}T_{14}} \nn\\
&\qquad +\sqrt{T_{23}T_{34}T_{24}}  + \sqrt{T_{12}T_{23}T_{34}T_{14}} \nn \\
&\qquad + \sqrt{T_{12}T_{13}T_{24}T_{34}} + \sqrt{T_{13}T_{14}T_{23}T_{24}} 
\end{align}
for the transition probabilities of any 4 pure states.  

Note that if $|\psi_4\rangle$ is orthogonal to the first three states, i.e, $T_{j4}=0$ for $j=1,2,3$, then Eq.~(\ref{our3}) reduces to the OUR in Eq.~(6) of the main text for any 3 pure states.  Since the latter can be saturated by suitable qubit states, it follows that the above OUR can be saturated by suitable qutrit states.

\end{document}